\documentclass[12pt]{article}
\usepackage{epsfig,subfigure,color,amsmath}
\usepackage{authordate1-4}

\begin{document}
\title{A climatic thermostat making Earth habitable}

\author{Peter D. Ditlevsen, \\
The Niels Bohr Institute, Department of Geophysics,\\
University of Copenhagen, Juliane Maries Vej 30,\\
 DK-2100 Copenhagen O, Denmark.\\
}

\date{\today}
\maketitle
{\bf
The mean surface temperature on Earth and other planets with atmospheres is determined by the radiative balance between the non-reflected incoming solar
radiation and the outgoing long-wave black-body radiation from the atmosphere.
The surface temperature is higher than the black-body temperature due to the
greenhouse warming. Balancing the ice-albedo cooling and the greenhouse warming 
gives rise to two stable climate states. A cold climate state with a completely ice-covered planet, called Snowball Earth, and a warm state similar to our 
present climate where greenhouse warming prevents the total glacition. The warm
state has dominated Earth in most of its geological history despite a 30 \% fainter young Sun. The warming could have been controlled by a greenhouse thermostat operating by temperature control of the weathering process depleting the atmosphere from $CO_2$. This temperature control has permitted life to evolve as early as the end of the heavy bombartment 4 billion years ago. 
}

\section*{Introduction}
Primitive life has existed on Earth since early in its geological history. Stromatolites, which are thought to be fosils of photosythesizing cyanobacteria are found as old as 3.5 Ga (giga-anni = billion years) \cite{schorpf:1993}. Isotope fractionation in carbon found in 3.8 Ga old rocks form Isua, Greenland indicates a biological origin \cite{rosing:1999}. This is perhaps within a few hundred million years the earlies time of stable planetary climate possible for life. In the prior heavy-bombardment period impacts releasing energies enough to evaporate the entire ocean would probably have sterilized the Earth if life existed at that time \cite{kasting:2003}.    

At some time around the Archaean-Proterozoic border, 2.5 Ga BP, the atmospheric content of oxygen began rising to its present level, making way for dominance of aerobic life forms. For a very long period after then there is no evidence of biological evolution. The eukaryotic (cells with a nucleus) and multicellular life seems to have evolved as late as 0.6 Ga BP at the Cambrian explosion, or perhaps in the late Precambrium where the soft body Ediacara fauna originates. This could indicate that the oxygen level rose slowly, reaching levels needed for effective oxygen transports and metabolisms in multicellular organisms at the Cambrian explosion.

The habitable zone (HZ) around a sun-like star is defined by the requirement of liquid water at the surface of planets or moons in the zone. This zone is primarily determined by the radiative output from the star. The dependence of the radiative flux, which goes as inverse
distance squared makes the HZ relatively narrow. At the present time our solar system contains only Earth within the HZ \cite{kasting:2003}. 

The long existence of life indicates that the surface temperature on Earth has indeed kept within the narrow range permitting liquid water for surprisingly long, despite a 30 \% lower solar flux at the beginning of Earth's history. This apparent paradox is dobted "The Faint Young Sun Problem" \cite{kasting:1988}. 
The solution to the problem could be a selfregulatory mechanism, either through a biotic feedback \cite{lovelock:1982} or a geochemical regulation \cite{owen:1979,walker:1981}. The former can only be at play after the evolution and expansion of life to the planetary scale, and can therefore not explain the
stable climate conditions necessary for the initiation of life.    
      
A geochemical regulation could, besides extending the time for habitability of a planet, potentially widen the HZ. There is now mouting evidence that Mars had
liquid water on the surface in its early history and was thus within the HZ then 
despite the lower solar luminosity. 

\section*{The radiative energy balance}
The surface temperature of the planet is determined from the balance between incoming solar radiation $R_i$ and outgoing black-body radiation $R_o$.
This energy balance depends, among other factors, on the planetary albedo.
The albedo of an object is the fraction of the sunlight hitting the object which is reflected.
The planetary albedo is not a constant factor; it depends, through the amount of clouds and ice, on the state of the climate itself. The feedback of clouds on temperature is very complicated. It depends on the height in the atmosphere where the
clouds are and the state of
the atmosphere surrounding the clouds. The clouds cool by reflecting the incoming radiation and they heat by trapping the
outgoing radiation. Ice and snow on the surface
unambiguously cool by
reflecting the incoming short-wave radiation, so the amount of ice and snow influences the planetary
albedo.
This effect can be described in a model of the climate represented by just one parameter, the 
mean surface temperature T \cite{crafoord:1978,ghil:1985}. This temperature determines the long-wave radiation and the reflection of the
short-wave
radiation through the albedo.
The amount of ice and snow is larger when the temperature is lower so the
lower the temperature the higher albedo.
If the temperature is below some low temperature $T_1$ the planet will be completely ice covered
and a
further decrease in temperature cannot increase the albedo above the value $\alpha_1$
for the fully ice covered planet. If the temperature is above some other
high temperature
$T_2$ the ice is completely melted and a further increase in temperature will not lead to a decrease
in albedo below the value $\alpha_2$ for the ice free planet. The
simplest functional form is a linear dependence of the albedo on temperature in between these two
temperatures. This is
a reasonable choice when no other information is available a priory. We then have the
relation
\begin{equation}
\alpha(T)=\alpha_1 {\cal I}_{[0,T_1]}+\frac{(T_2-T)\alpha_1+(T-T_1)\alpha_2}{T_2-T_1}
{\cal I}_{(T_1,T_2]}+\alpha_2{\cal I}_{(T_2,\infty)}
%\begin{cases}
%\alpha_1& T\le T_1 \\
%\frac{(T_2-T)\alpha_1+(T-T_1)\alpha_2}{T_2-T_1}& T_1<T\le T_2 \\
%\alpha_2& T >T_2
%\end{cases}
\end{equation}
where ${\cal I}_{(a,b]}$ is the indicator function for the interval $(a,b]$.
The change of the temperature $T$ is determined by the difference
$R_{\text{i}}-R_{\text{o}}$ in incoming and outgoing
radiation according to the equation
\begin{equation}
c\frac{\text{d} T}{\text{d} t}= R_{\text{i}}-R_{\text{o}}=[1-\alpha(T)]S-\sigma g(T) T^4
\label{bs}
\end{equation}
where $c$ is the heat capacity, $\sigma$ is the Stefan-Boltzmann constant and $S=\tilde{S}/4$ is a quarter of the solar constant.
(The quarter comes from the
ratio of the cross-sectional area to the surface area of the sphere).
The factor $g(T)$ expresses the atmospheric greenhouse effect. 
The black-body temperature $T_{\text{\small b-b}}$ of the planet is the temperature at the level in the atmosphere
from where the long-wave radiation is emitted. This level is the height of the optical thickness at the long-wave band 
seen from space. Depending on greenhouse gasses and clouds the level of outgoing radiation is approximately 3 kilometers 
above the surface. The difference between the black-body temperature and the surface temperature is the greenhouse 
warming (or cooling). On Earth the atmosphere is transparent to the sunlight which thus heats the surface, this, in 
turn, heats the atmosphere from below. The lower atmosphere (the troposhere)  thus experiences a negative lapserate 
(temperature change with height). The lapserate depends in a complicated way on the static stability and atmospheric 
dynamics. In the present climatic conditions the lapserate is of the order -10K/km, thus the greenhouse effect on 
Earth is approximately 3km$\times$10K/km=30K. Without the 
greenhouse effect there would be no liquid water at the surface of the Earth.   
The atmospheric greenhouse
effect, the change in cloudiness and other factors must all be expressed through
the "transfer function" $g(T)$, where $T$ represents a mean surface temperature, from which the black-body temperature 
is derived; $\sigma g(T)T^4=\sigma T_{\text{\small b-b}}^4$. 
The simple model (\ref{bs}) is relevant because the 
behavior does not depend critically on the specific choice of the
parameters \cite{budyko:1969,sellers:1969,crafoord:1978}.

\begin{figure}[!H]
\begin{center}\epsfxsize=12cm %\textwidth
\epsffile{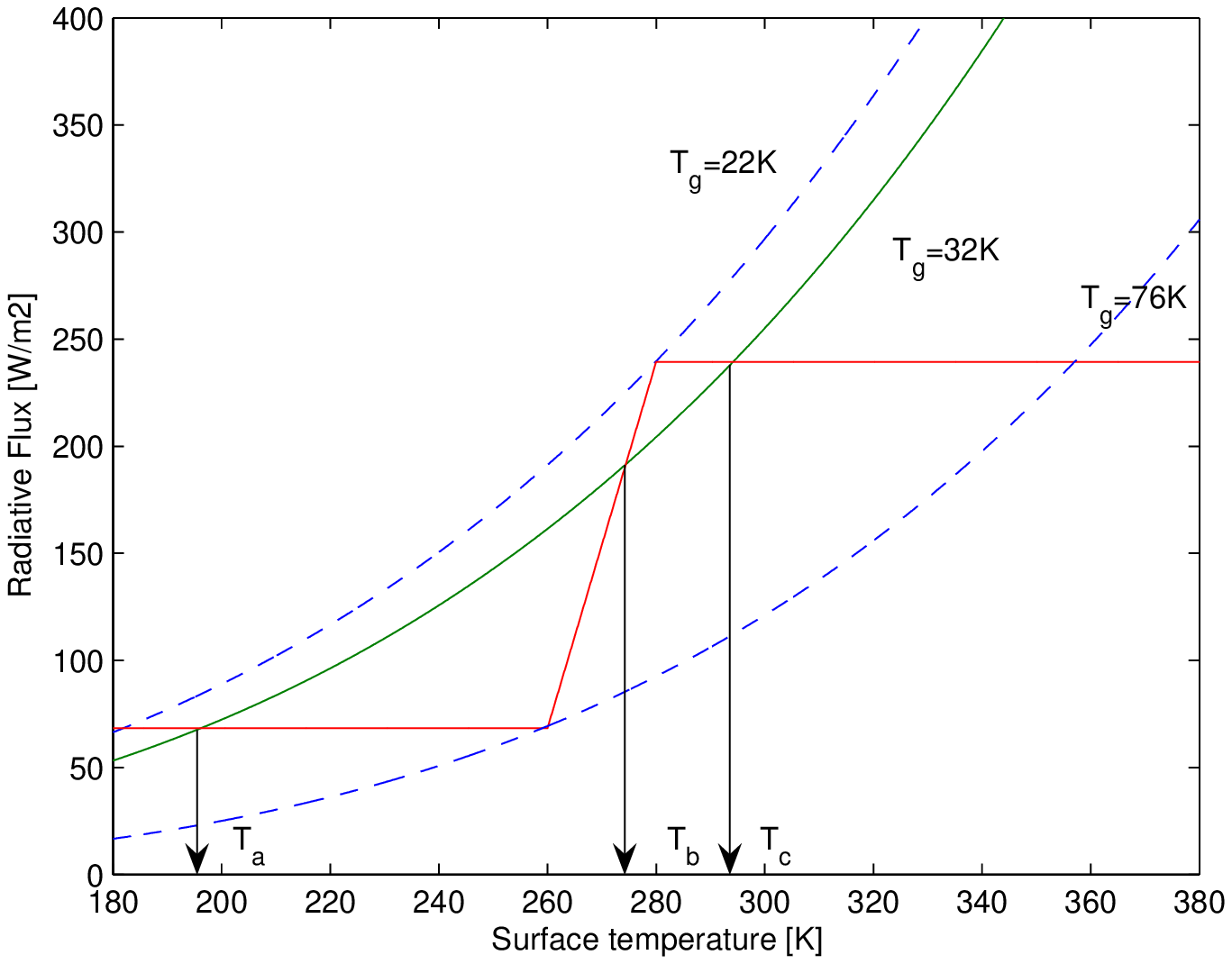}\caption[]{}
\label{fig1}
\end{center}\end{figure}

The behavior of (\ref{bs}) is easily understood from a graphic representation. Figure \ref{fig1}
shows the incoming and
outgoing radiations as a function of temperature.
There are three temperatures $T_a,T_b,T_c$ for which the curves cross
such that the incoming and outgoing radiations are in balance. These points, the fixed points, are the stationary solutions to
(\ref{bs}). Consider
the climate to be at point $T_a$. If some small perturbation makes the temperature become lower than
$T_a$ we will have
$R_i>R_o$ implying that $cdT/dt >0$ and the temperature will rise to $T_a$ (see figure 1). If, on the other hand, the
perturbation is positive
and the temperature is a small amount larger than $T_a$ we have $R_i<R_o $ implying that $cdT/dt<0$
and the
temperature will decrease to $T_a$ again. Thus $T_a$ is a stable fixed point. 
A similar analysis shows that $T_b$ is an unstable fixed
point and $T_c$ is a stable fixed point.
If the temperature at some initial time is lower than
$T_b$ it will eventually reach the temperature $T_a$ and if it is higher than $T_b$ it will reach the temperature $T_c$.
The present climate is the climate state $T_c$ where the ice albedo does not play a significant role in cooling the Earth.

The outgoing radiation depends on the surface temperature $T$ through the atmospheric concentration of greenhouse 
gasses and clouds. Expressing the greenhouse effect in terms of the difference between the the surface temperature and the
black-body temperature $T_g=(T-T_{\text{\small{b-b}}})$ the outgoing radiation 
may be written $g(T) \sigma T^4 = \sigma (T-T_g-0.3*[T-270])^4 =R_o(T,T_g)$. The last factor in the paranthesis is an
empirical factor expressing the increase in greenhouse effect with temperature due to increase of atmospheric water vapor
with temperature. 
%The form of $R_o(T,T_g)$ is very close to the one previously found (\cite{sellers:1969}).   

\begin{figure}[!H]
\begin{center}\epsfxsize=12cm %\textwidth
\epsffile{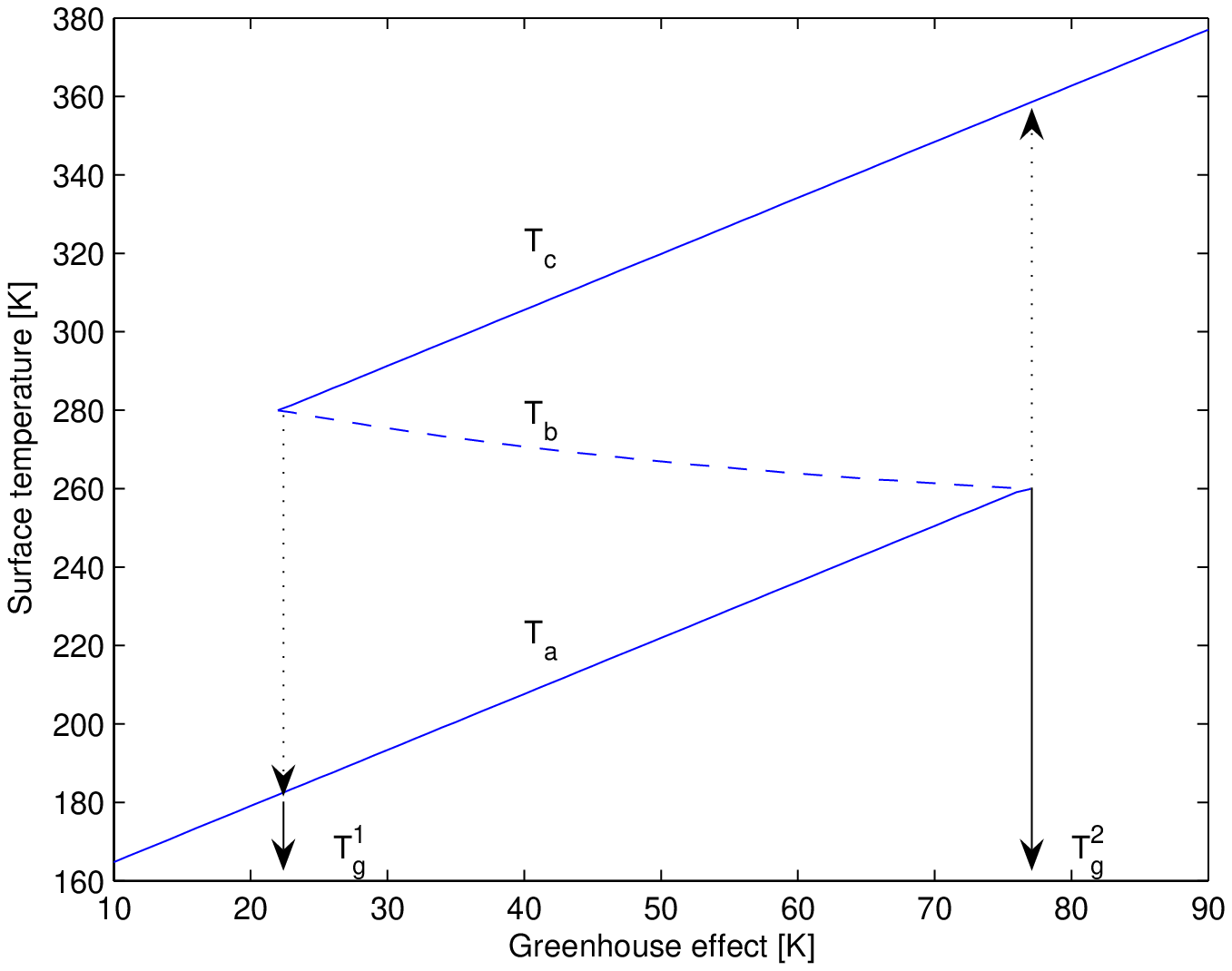}\caption[]{}
\label{fig2}
\end{center}\end{figure}

Consider the climate state represented by $T_c$ in the situation $T_g < T_g[\text{Present day}]$. Then for $T_g$ not
too small,
corresponding to the upper dashed blue curve in figure \ref{fig1}, the equilibrium temperature is lowered a
little, as we would expect
when the greenhouse warming decreases. However, if $T_g$ becomes smaller than some value $T_g^{(1)}$ the
two curves do not
cross in more that one point and there is no stable fixed point near $T_c$. The
climate will then run
into the only stable fixed point $T_a$ which is still present. A saddle-node bifurcation has occurred resulting in a large change of
climate. If $T_g$
grows again the
climate state $T_c$ will not recover until $T_g$ exceeds some other value $T_g^{(2)}>T_g[\text{Present day}]$ and the system 
returns through a hysteresis loop.
For each value of $T_g$ we have either one or three fixed points and we can plot the fixed points in a bifurcation diagram as functions of $T_g$ (figure \ref{fig2}).
The two full curves represent the stable fixed
points and the middle dashed curve represents the unstable fixed point.
The unstable and one of the stable points coincide at the
bifurcation points $T_g^{(1)}$ and $T_g^{(2)}$.

The
stable climate state $T_a$ corresponds to a totally ice covered planet. 
The totally ice covered planet has been termed
"Snowball Earth" \cite{hoffman:1998}. There is geological evidence of such an extreme ``deep freeze" climate
several times in the late Neoproterozoic period around 0.7 Ga BP. This is based
on findings of glacial deposits like moraine in many places which at those times were
near the equator. The speculated way out of the deep freeze is the following: The balance in
geological timescales between silicate weathering, binding atmospheric $CO_2$ into rocks and
volcanic out-gassing of $CO_2$ was changed during the deep freeze. Due to the cold conditions
the atmosphere
dried out and weathering effectively stopped. Unchanged volcanic out-gassing resulted in an ever increasing
amount of $CO_2$ in the atmosphere. At some point, after about 80 million years, this would
result in a greenhouse warming strong enough to melt the ice. The warming would then be
almost explosive with global mean temperature going from some $-40^\circ C$ to
some $+50^\circ C$ within a few years. 
This kind of dramatic climatic changes will strongly stress the planetary biota. 

\section*{Stability of the climatic state}
The planetary climate is influenced by internal and external factors which can push the
climate state away from the equilibrium position. If the perturbations are small the 
equilibrium state will be restored within a typical timescale, depending on the size
of the perturbation and the strength of the restoring force.   
The fluctuations can be represented as an independent white noise $\eta(t)$ with intensity
$\tilde{\sigma}$. Linearizing (\ref{bs}) around the equilibrium state $T_c$ we get:

\begin{equation}
c\frac{\text{d}T}{\text{d}t}=-\alpha(T-T_c)+\tilde{\sigma} \eta.
\label{ou}
\end{equation}
The parameter $-\alpha$ is the expansion coefficient for the right hand side of (\ref{bs}).
Equation (\ref{ou}) is the Ohrnstein-Uhlenbeck process \cite{gardiner:1985}, where the variance is
$\langle(T-T_c)^2\rangle=\tilde{\sigma}^2/2\alpha$ (see figure 3) and the timescale for restoring the equilibrium temperature can
be defined as the autocorrelation time  $\tau=\alpha^{-1}$. The stability against the random
perturbations is thus measured by $\alpha$, the larger $\alpha$ the smaller is the response 
to the "noise" and the faster the perturbation is forgotten. The stability of the climate state
represented by the temperature $T_c$ governs the conditions for biota. However, a long timescale 
stability is not ensured if the parameters governing the value of $T_c$ itself is changing. 
The early sun was about 30 \% less luminous than today, which implies that $T_c$ determined by
(\ref{bs}) would be lower than today. 
For a fixed value of the greenhouse gas mixing ratio the 
equilibrium temperature can be found from (\ref{bs}) as a function of the solar flux $S$. 
This is shown schematically in figure \ref{fig4}. At times prior to approximately 2 Ga before
present the solar flux was lower and        
not permitting liquid water at the surface of the Earth, 
contrary to what is observed, except for the spells of Snowball Earth in the 
late Neoproterozoic period (0.7-0.6 Ga BP). A possible solution to this enigma is that the concentration of 
greenhouse gasses was much higher in the early atmosphere. 

\section*{The greenhouse thermostat}
The atmospheric concentration of $CO_2$
in geological timescales depends on the balance between outgassing through volcanos and burrial
through weathering of silicate rocks. The rates of change of atmospheric $CO_2$ concentration 
governed by these two factors does not depend on the concentration
itself. This implies that they will not balance unless they are exactly equal and opposite. The outgassing 
is independent of the surface temperature while the silicate
weathering rate is stongly depending on temperature. Weathering requires precipitation dissolving atmospheric 
$CO_2$ in the form of carbonic acid. The rate of precipitation is stongly temperature dependent. In the present climatic 
conditions weathering takes place in the tropics while it is almost absent in the dry and cold polar regions. 
The simplest way to describe the effect of temperature on silicate weathering is by a step function;

\begin{equation}
\frac{\text{d}[CO_2]}{\text{d}t}=F_o-F_w \theta(T-T_0)
\end{equation}
where $F_o$ is the rate of outgassing, $F_w$ is the rate of weathering, $\theta(T-T_0)$ is the heaviside step 
function and $T_0=285K$ is the temperature below which the is no weathering \cite{walker:1981}. If $F_w>F_o$ the $CO_2$ will be depleted 
from the atmosphere. This will, however, cool the planet by diminishing the greenhouse effect. When the
temperature falls to $T_0$ weathering stops and the atmospheric $CO_2$ rises again by outgassing. In terms of the energy 
balance the greenhouse factor $T_g$ should represent the part of the greenhouse which is independent of time 
while the $CO_2$ greenhouse effect must be described as an additional factor. The energy balance (\ref{bs}) then becomes:

\begin{equation}
c\frac{\text{d}T}{\text{d}t}=[1-\alpha(T)]S-\sigma (T-T_g-0.3[T-270])^4+f\theta(T_0-T)
\end{equation}  
     
where $f\theta(T_0-T)$ is the additional $CO_2$ greenhouse heating.
The behavior is now fundamentally different from the behavior of (\ref{bs}).
For $T_c<T_0$ the warm climate state is given by
\begin{equation}
T_c^<= [\{1-\alpha_2)S+f/\sigma\}^{1/4}+T_g-51]/0.8
\end{equation}
If $T_c^< >T_0$  we have
\begin{equation}
T_c^>=[\{1-\alpha_2)S/\sigma\}^{1/4}+T_g-51]/0.8
\end{equation}
But $T_c^> < T_c^< \Rightarrow T_c^< <T_0$ contrary to the assumption. In this case
$T_c$ is not a steady state solution. The surface temperature will increase until
it reaches $T_0$ where weathering efficiently depletes the atmosphere from $CO_2$ and
the temperature drops again. This mechanism is a greenhouse thermostat.
The weaker heating from the faint young sun is thus automatically compensated by a stronger early $CO_2$ greenhouse.      
The resulting constant surface temperature is indicated as the fat blue curve in figure 4.
In the future the thermostat will not be functional since the atmosphere will be
depleted from $CO_2$ so the increasing solar flux will result in a steady increase
in surface temperature. 

\section*{Perspective}
The surface temperature on Earth has been relatively stable permitting liquid water necessary for biology as we know it in most of its geological history. This is despite a relatively large change in solar flux and atmospheric content of greenhouse gasses which suggests a thermostat mechanism controlling the surface temperature. A possible thermostat could be operational through silicate weathering depleting the atmosphere from the constantly outgassed volcanic $CO_2$. The temperature at which weathering becomes active will determine the surface temperature by adjusting the atmospheric greenhouse $CO_2$ level for the long-wave outgoing radiation balancing the incoming solar radiation. The thermostat will regulate temperature until the solar flux is so large that it can maintain a surface temperature above the temperature where weathering becomes active without the $CO_2$ greenhouse warming. From that point in time the surface temperature will increase with
the increasing solar flux. The greenhouse thermostat is keeping the climate warmer than the stable Snowball Earth situation where the ice albedo keeps the surface globally below the freezing point of water. Relatively constant temperatures over geological time determined by geochemical thermostats could be important for initiation of biological life and potentially widen the habitable zone around a sun-like star. When bacterial type life or the like is established on a planet it could provide a thermostat by itself, as suggested by the gaia hypothesis \cite{lovelock:1982}. Life on Earth is today a dominant player in controlling the atmospheric levels of $O_2$, $CO_2$ and $CH_4$.

\newpage
\bibliographystyle{authordate1}
\bibliography{/disk3/pditlev/documents/shell_model/full/climate}     

\newpage
\begin{center}
FIGURE CAPTIONS
\end{center}
\newcounter{fig}
\begin{list}{Fig. \arabic{fig}}
{\usecounter{fig}\setlength{\labelwidth}{2cm}\setlength{\labelsep}{3mm}}

\item
The incoming (red) and outgoing (blue and green) fluxes as a function of global 
temperature. When the greenhouse warming is below 22 K the warm stable climate
state at $T_c$ disappears through a saddle-node bifurcation and the 'deep freeze' state
at $T_a$ is the only stable climate state.   

\item
The bifurcation diagram for the energy balance. If the greenhouse warming fall below
$T_g^{(1)}$ the climate will fall into the Snowball Earth climate. 
The greenhouse warming has to
exceed $T_g^{(2)}$ for the planet to leave the Snowball Earth climate. The dodded
arrows indicate a hysteresis loop. 

\item
The stability of the warm climate state against random fluctuations is determined by
the radiative cooling feedback represented by the parameter $-\alpha$.

\item
The surface temperature has, except for a few Snowball Earth episodes in the
late Neoproterozoic, been above the freezing point of water for most of
Earths geological history since the heavy bombardment epoch. The surface temperature
has been governed by the greenhouse thermostat. The thermostat was operational
until the solar luminosity became strong enough perhaps 1.5-1 Ga BP. The present
atmospheric content of $CO_2$ and other gasses is regulated by biology itself, 
not included in the graph.   
\end{list}

\begin{figure}[!H]
\begin{center}\epsfxsize=12cm %\textwidth
\epsffile{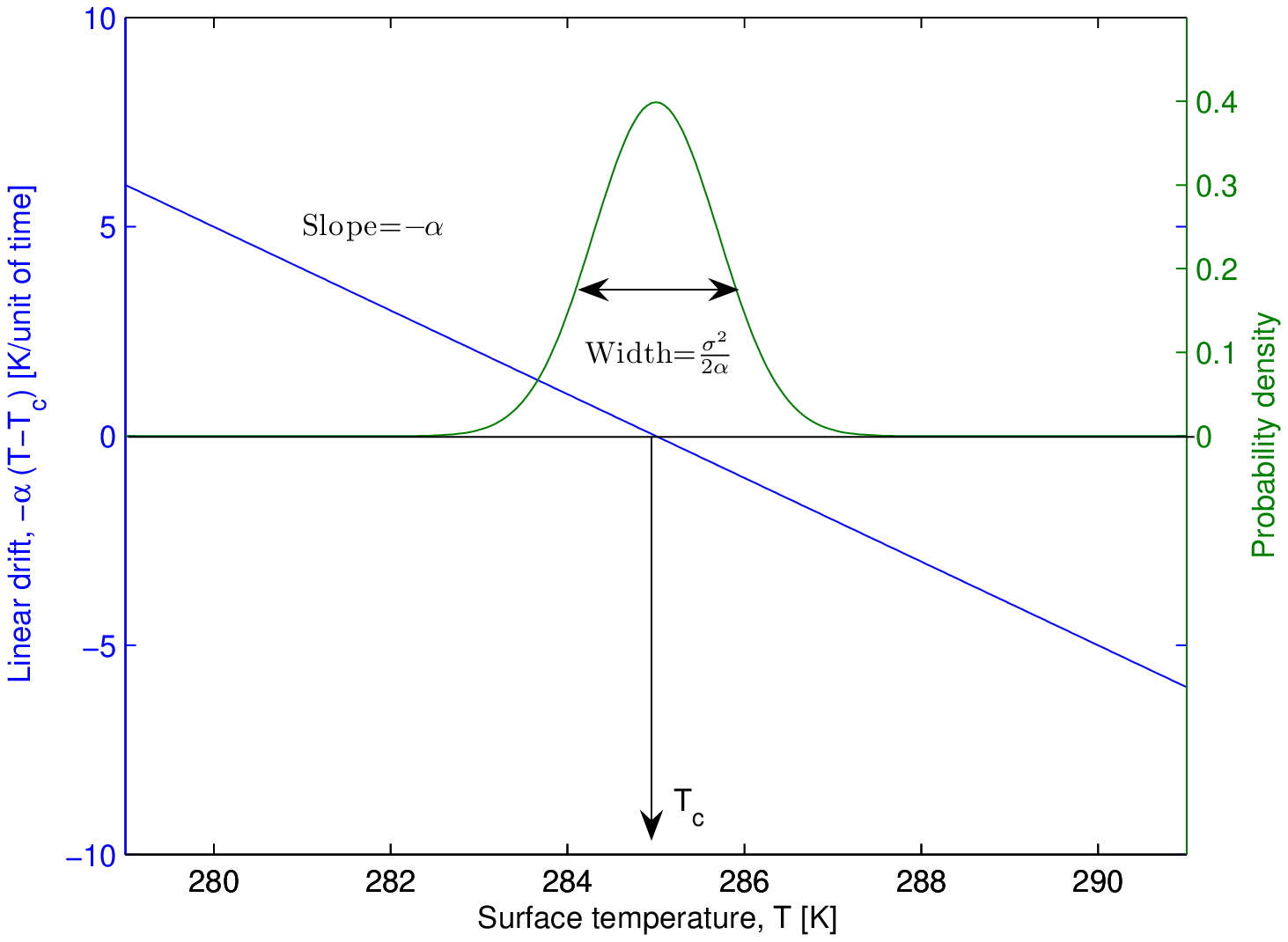}\caption[]{}
\label{fig3}
\end{center}\end{figure}

\begin{figure}[!H]
\begin{center}\epsfxsize=12cm %\textwidth
\epsffile{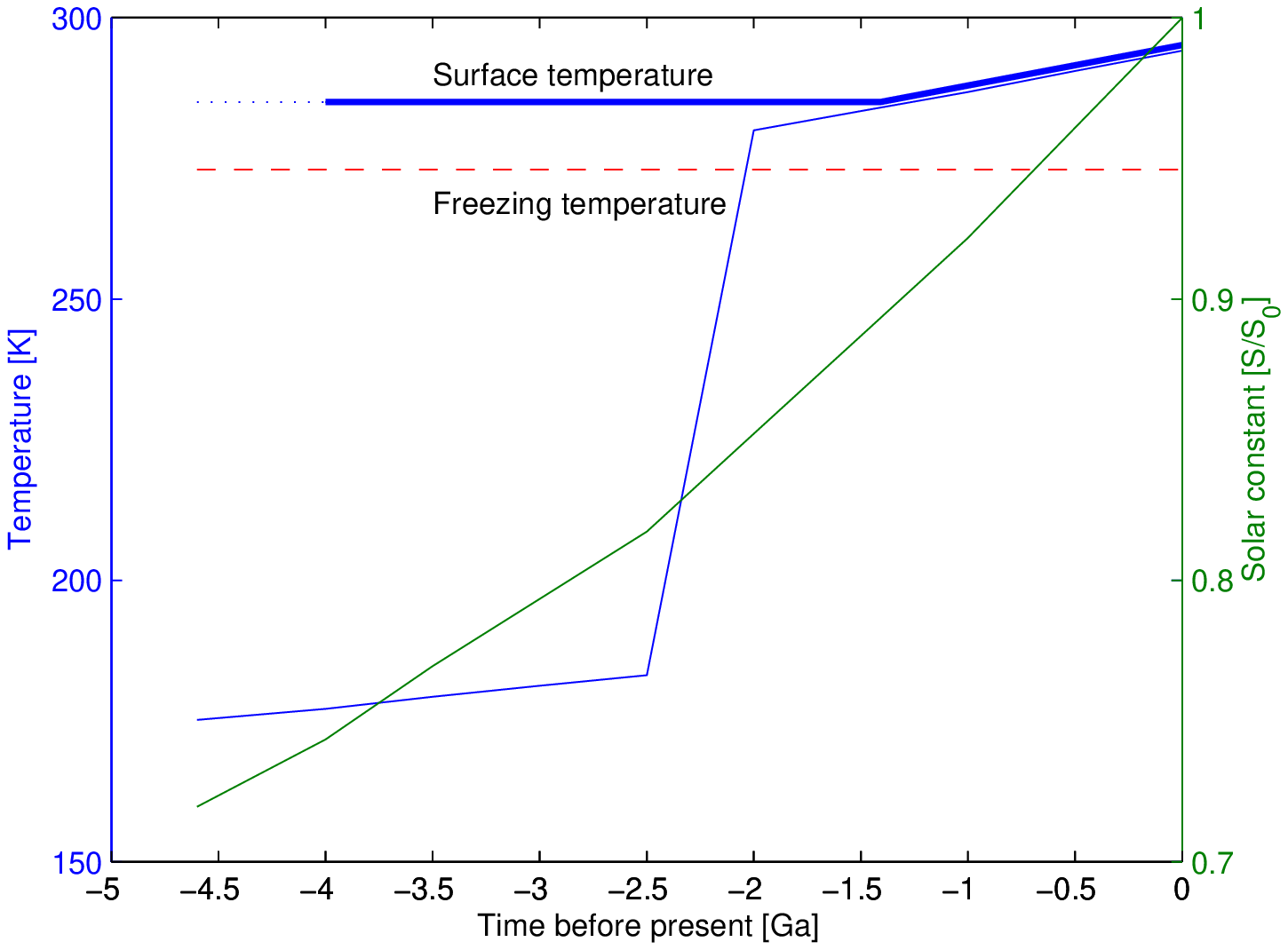}\caption[]{}
\label{fig4}
\end{center}\end{figure}

\end{document}